\documentstyle{article}

\newcommand{\mb}[1]{\mbox{\normalsize\boldmath $#1$}}
\newcommand{\GeV}{\,{\rm GeV}}
\catcode`\@=11
\def\marginnote#1{}
\newcount\hour
\newcount\minute
\newtoks\amorpm
\hour=\time\divide\hour by60
\minute=\time{\multiply\hour by60 \global\advance\minute by-\hour}
\edef\standardtime{{\ifnum\hour<12 \global\amorpm={am}%
        \else\global\amorpm={pm}\advance\hour by-12 \fi
        \ifnum\hour=0 \hour=12 \fi
        \number\hour:\ifnum\minute<10 0\fi\number\minute\the\amorpm}}
\edef\militarytime{\number\hour:\ifnum\minute<10 0\fi\number\minute}
\def\draftlabel#1{{\@bsphack\if@filesw {\let\thepage\relax
   \xdef\@gtempa{\write\@auxout{\string
      \newlabel{#1}{{\@currentlabel}{\thepage}}}}}\@gtempa
   \if@nobreak \ifvmode\nobreak\fi\fi\fi\@esphack}
        \gdef\@eqnlabel{#1}}
\def\@eqnlabel{}
\def\@vacuum{}
\def\draftmarginnote#1{\marginpar{\raggedright\scriptsize\tt#1}}
\def\draft{\oddsidemargin -.5truein
        \def\@oddfoot{\sl preliminary draft \hfil
        \rm\thepage\hfil\sl\today\quad\militarytime}
        \let\@evenfoot\@oddfoot \overfullrule 3pt
        \let\label=\draftlabel
        \let\marginnote=\draftmarginnote
   \def\@eqnnum{(\theequation)\rlap{\kern\marginparsep\tt\@eqnlabel}%
\global\let\@eqnlabel\@vacuum}  }
\def\numberbysection{\@addtoreset{equation}{section}
        \def\theequation{\thesection.\arabic{equation}}}
\def\underline#1{\relax\ifmmode\@@underline#1\else
        $\@@underline{\hbox{#1}}$\relax\fi}
\def\titlepage{\@restonecolfalse\if@twocolumn\@restonecoltrue\onecolumn
     \else \newpage \fi \thispagestyle{empty}\c@page\z@
        \def\thefootnote{\fnsymbol{footnote}} }
\def\endtitlepage{\if@restonecol\twocolumn \else \newpage \fi
        \def\thefootnote{\arabic{footnote}}
        \setcounter{footnote}{0}}  
\catcode`@=12
\relax
\def\figcap{\section*{Figure Captions\markboth
        {FIGURECAPTIONS}{FIGURECAPTIONS}}\list
        {Figure \arabic{enumi}:\hfill}{\settowidth\labelwidth{Figure 999:}
        \leftmargin\labelwidth
        \advance\leftmargin\labelsep\usecounter{enumi}}}
 \relax
\def\tablecap{\section*{Table Captions\markboth
        {TABLECAPTIONS}{TABLECAPTIONS}}\list
        {Table \arabic{enumi}:\hfill}{\settowidth\labelwidth{Table 999:}
        \leftmargin\labelwidth
        \advance\leftmargin\labelsep\usecounter{enumi}}}
 \relax
\def\reflist{\section*{References\markboth
        {REFLIST}{REFLIST}}\list
        {[\arabic{enumi}]\hfill}{\settowidth\labelwidth{[999]}
        \leftmargin\labelwidth
        \advance\leftmargin\labelsep\usecounter{enumi}}}
 \relax
\makeatletter
\newcounter{pubctr}
\def\publist{\@ifnextchar[{\@publist}{\@@publist}}
\def\@publist[#1]{\list
        {[\arabic{pubctr}]\hfill}{\settowidth\labelwidth{[999]}
        \leftmargin\labelwidth
        \advance\leftmargin\labelsep
        \@nmbrlisttrue\def\@listctr{pubctr}
        \setcounter{pubctr}{#1}\addtocounter{pubctr}{-1}}}
\def\@@publist{\list
        {[\arabic{pubctr}]\hfill}{\settowidth\labelwidth{[999]}
        \leftmargin\labelwidth
        \advance\leftmargin\labelsep
        \@nmbrlisttrue\def\@listctr{pubctr}}}
 \relax
\makeatother
\catcode`\@=11
\def\section{\@startsection {section}{1}{0pt}{-3.5ex plus -1ex minus
 -.2ex}{2.3ex plus .2ex}{\raggedright\large\bf}}
\catcode`\@=12
\newskip\humongous \humongous=0pt plus 1000pt minus 1000pt
\def\caja{\mathsurround=0pt}

\newif\ifdtup
\def\panorama{\global\dtuptrue \openup1\jot \caja
        \everycr{\noalign{\ifdtup \global\dtupfalse
        \vskip-\lineskiplimit \vskip\normallineskiplimit
        \else \penalty\interdisplaylinepenalty \fi}}}
\def\eqalignno#1{\panorama \tabskip=\humongous
        \halign to\displaywidth{\hfil$\displaystyle{##}$
        \tabskip=0pt&$\displaystyle{{}##}$\hfil
        \tabskip=\humongous&\llap{$##$}\tabskip=0pt
        \crcr#1\crcr}}

\def\oldreffmt#1{\rlap{[#1]} \hbox to 2\parindent{}}

\def\figfmt#1{\rlap{Figure {#1}} \hbox to 1in{}}

\def\abs#1{\left| #1\right|}

\def\ltap{\raisebox{-.4ex}{\rlap{$\sim$}} \raisebox{.4ex}{$<$}}

\catcode`\@=11
\def\eqnarray{\stepcounter{equation}\let\@currentlabel=\theequation
\global\@eqnswtrue
\global\@eqcnt\z@\tabskip\@centering\let\\=\@eqncr
\gdef\@@fix{}\def\eqno##1{\gdef\@@fix{##1}}%
$$\halign to \displaywidth\bgroup\@eqnsel\hskip\@centering
  $\displaystyle\tabskip\z@{##}$&\global\@eqcnt\@ne
  \hskip 2\arraycolsep \hfil${##}$\hfil
  &\global\@eqcnt\tw@ \hskip 2\arraycolsep $\displaystyle\tabskip\z@{##}$\hfil
   \tabskip\@centering&\llap{##}\tabskip\z@\cr}

\def\@@eqncr{\let\@tempa\relax
    \ifcase\@eqcnt \def\@tempa{& & &}\or \def\@tempa{& &}
      \else \def\@tempa{&}\fi
     \@tempa \if@eqnsw\@eqnnum\stepcounter{equation}\else\@@fix\gdef\@@fix{}\fi
     \global\@eqnswtrue\global\@eqcnt\z@\cr}

\catcode`\@=12
\font\tenbifull=cmmib10 
\font\tenbimed=cmmib10 scaled 800
\font\tenbismall=cmmib10 scaled 666
\relax

\def\thefootnote{\fnsymbol{footnote}}
\begin{document}
\begin{titlepage}
\begin{center}\today
\hfill LBL-36907\\
\hfill UCB-PTH-95/06\\
\hfill IFUP -- TH 13/95\\
\hfill hep-ph/9504373\\

\vskip .5in

{\huge \bf Hadronic Flavor and CP Violating\\
Signals of Superunification\footnote{This work was
supported in part by the Director, Office of Energy Research,
Office of High Energy and Nuclear Physics, Division of High
Energy Physics of the U.S. Department of Energy under Contract
DE-AC03-76SF00098 and in part by the National Science
Foundation under grant PHY-90-21139.}}

\vglue 1.5cm
{\large\bf Riccardo Barbieri$^\dagger$, Lawrence Hall$^\ddagger$ \rm and
       \bf Alessandro Strumia$^\dagger$\\}
\vglue 0.8cm
$\dagger$~{\em Dipartimento di Fisica, Universit\`a di Pisa \& \\}
{\em INFN, Sezione di Pisa, I-56126 Pisa, Italy\\}
\vglue 0.4cm
$\ddagger$~{\em Theoretical Physics Group, Lawrence Berkeley Laboratory\\
and Department of Physics, University of California, Berkeley, California
94720}
\vfill\large{\bf Abstract}
\end{center}

The flavor changing and CP violating phenomena predicted in supersymmetric
unified theories as a consequence of the large top quark Yukawa coupling,
are investigated in the quark sector and compared with related phenomena
in the lepton sector, considered previously.
In particular we study $\varepsilon_K$, $\varepsilon_K'/\varepsilon_K$,
$\Delta m_B$, $b\to s\gamma$, the neutron electric dipole moment, $d_n$, and~CP
violation in neutral $B$ meson decays, both in minimal~SU(5) and~SO(10)
theories.
The leptonic signals are generically shown to provide more significant tests of
quark-lepton unification. Nevertheless, mostly in the~SO(10) case, a variety of
hadronic signals is also possible, with interesting correlations among them.
\normalsize

\vfill~\vfill~
\end{titlepage}

\renewcommand{\thepage}{\roman{page}}
\thispagestyle{empty}
\setcounter{page}{2}
\mbox{ }
\vskip 1in
\begin{center}
{\bf Disclaimer}
\end{center}

\vskip .2in

\begin{scriptsize}
\begin{quotation}
This document was prepared as an account of work sponsored by the United
States Government. While this document is believed to contain correct
 information, neither the United States Government nor any agency
thereof, nor The Regents of the University of California, nor any of their
employees, makes any warranty, express or implied, or assumes any legal
liability or responsibility for the accuracy, completeness, or usefulness
of any information, apparatus, product, or process disclosed, or represents
that its use would not infringe privately owned rights.  Reference herein
to any specific commercial products process, or service by its trade name,
trademark, manufacturer, or otherwise, does not necessarily constitute or
imply its endorsement, recommendation, or favoring by the United States
Government or any agency thereof, or The Regents of the University of
California.  The views and opinions of authors expressed herein do not
necessarily state or reflect those of the United States Government or any
agency thereof or The Regents of the University of California and shall
not be used for advertising or product endorsement purposes.
\end{quotation}
\end{scriptsize}

\vskip 2in

\begin{center}
\begin{small}
{\it Lawrence Berkeley Laboratory is an equal opportunity employer.}
\end{small}
\end{center}

\newpage
\renewcommand{\thepage}{\arabic{page}}
\setcounter{page}{1}
\section{Introduction}\small
The most widely discussed signatures of grand unification, studied since the
1970's, are proton decay, neutrino masses, fermion mass relations and the weak
mixing angle prediction.
The precise measurement of the weak mixing angle at $Z$ factories suggests
that these theories should incorporate weak-scale supersymmetry,
making superpartner mass relations a further signature.
In recent papers we have identified new signatures for supersymmetric
unification, with supersymmetry broken as in supergravity,
which provide signals which are less model dependent than those of proton
decay,
neutrino masses and fermion mass relations.
These new signatures include lepton flavor violation~\cite{BH} and electric
dipole moments for the electron, $d_e$,
and for the neutron, $d_n$~\cite{DH}.
In a  detailed study of the lepton signals~\cite{BHS}, rates for
$\mu\to e\gamma$
and for $\mu\to e$ conversion in atoms and values for $d_e$ have been given
over the entire range of
parameter space of simple~SU(5) and~SO(10) models.
Further searches for these signals can probe selectron mass ranges
of $100 \div 200\GeV$ for~SU(5) and $300 \div 600\GeV$ for~SO(10),
and are clearly very powerful.

This new class of signals arises because the top Yukawa coupling
of the unified theory leads to very large radiative corrections to
the masses of those superpartners which are unified with the top.
In the lepton sector this leads to an
important non-degeneracy of the sleptons, giving lepton flavor mixing matrices
at neutral
gaugino vertices.
It is clear that this phenomena is not limited to the lepton
sector, and the purpose of this paper is to study the flavor changing and
CP violating phenomena induced by this mechanism
in the quark sector.
In particular we study $\varepsilon_K$,
$\varepsilon'_K/\varepsilon_K$, $\Delta m_B$,
$b \to s\gamma$, $d_n$ and~CP violation in neutral $B$ meson decay. We address
the following questions:
\begin{itemize}
\item[(A)] How strong a limit is placed on the parameter space of unified
models by present measurements of hadronic flavor and CP violation?
\item[(B)] Can future measurements of hadronic flavor and CP violation
provide a test of supersymmetric unification?
\item[(C)] If so, how does the power of these probes compare
with the lepton signals?
\end{itemize}
The answers to these questions are crucial in determining the optimal
experimental strategy for using this new class of signatures to probe
 unified theories.
For example, it is crucial to know whether  new
gluino-mediated contributions to $\varepsilon_K$ are so large that the
resulting
constraints on the parameter space preclude values of $\Gamma(\mu\to e\gamma)$
and $d_e$ which are accessible to future experiments.

If gluino-mediated flavor changing effects are found to be very large,
what are the best experimental signatures?
Three possibilities are:
\begin{itemize}
\item[i)] A pattern of CP violation in neutral $B$ meson decays which conflicts
with the prediction of the SM.
\item[ii)] Predictions for $\varepsilon_K$ and $\Delta m_B$ which deviate from
SM predictions for measured values of $m_t$ and~$V_{ub}$.
\item[iii)] A prediction for $B_s$ meson mixing $(x_s/x_d)$ which differs from
the prediction of the SM.
\end{itemize}
In Section~2 we define the minimal~SU(5) and~SO(10) models. The superpartner
spectrum for these models is discussed in Section~3. In Section~4 both analytic
and numerical results are given for the hadronic processes of interest in the
minimal~SO(10) model. We illustrate why in the~SU(5) case the hadronic signals
are less relevant. A study of these results, and a comparison with the
predictions for the lepton signals, allows us to answer questions~(A), (B) and
(C)  above. We aim at an overall view rather than at a detailed analysis of the
various effects. In Section~5  we mention aspects of the assumptions which
underlie our signatures. Our results are summarized in Section 6, where we also
show that our conclusions are not specific to the minimal models, but are more
generally true.

\section{The Minimal Models}
In this  paper we give results for flavor-changing and CP violating processes
in
two minimal superunified  models, one based on~SU(5) and the other on~SO(10).
The flavor structure of the models is constructed to be particularly simple,
and
the corresponding flavor mixing matrices of the low energy supersymmetric
theory
possess a very simple form, which directly reflects the unified group.
Nature is likely to be more complicated.
In the conclusions we discuss the extent to which our results are expected
to hold in more general models.
The predictions of the minimal models provide a useful reference point.
They provide a clean estimate of the size of the effects to be expected from
the top Yukawa coupling in theories where the top quark is unified with other
particles of the the third generation.
There are many additional flavor and CP violating effects which could be
generated from other interactions of the unified theory and could be much
larger
then those considered here.
While cancellations between different contributions can never be excluded, the
contribution given here provides a fair representation of the minimal amount to
be expected.
Circumstances which could lead to a significant reduction of the signals are
discussed in Section 5.

A crucial assumption, discussed in detail in Section 5, is that the
supersymmetry breaking is communicated to the fields of the Minimal
Supersymmetric Standard Model (MSSM) at a scale above the unification mass,
$M_{\rm G}$. For the analysis of this paper we assume the communication occurs
at the reduced Planck scale, $M_{\rm Pl}$, as in supergravity
theories~\cite{SuGra}, and furthermore we assume that at this scale the
supersymmetry breaking is universal.
This means that all scalars acquire a common supersymmetry break mass, $m^2_0$
and all trilinear superpotential interactions generate a supersymmetry breaking
trilinear scalar interaction with common strength given by the parameter $A_0$.
Similarly, there is a common gaugino mass $M_0$.
This boundary condition is not crucial to our effect; it is the simplest which
involves no flavor violation, so we can be sure that the signals we calculate
originate only from radiative effects of the top quark Yukawa coupling.

Before introducing the two minimal unified models, we review the flavor and CP
violating signals induced by the top quark Yukawa coupling of the
MSSM~\cite{5,6,BiGi}. The universal boundary condition on the supersymmetry
breaking interactions leads to the conservation of individual lepton numbers in
the MSSM, so we discuss only
the quark sector, where the superpotential can be written as:
$$
W_{\rm MSSM} = Q\overline{\mb{\lambda}}_U U^cH_2 + Q\mb{\lambda}_D D^c
H_1\eqno(1)
$$
where $\mb{\lambda}_D = \mb{V}^*\overline{\mb{\lambda}}_D, \mb{V}$ is the
Kobayashi-Maskawa (KM)
matrix, and $\overline{\mb{\lambda}}_U$ and $\overline{\mb{\lambda}}_D$
are real and diagonal Yukawa coupling matrices.
Throughout this paper we assume that the largest eigenvalue of
 $\mb{\lambda}_D$, $\lambda_b$, is sufficiently small that the only Yukawa
coupling which need be kept in the renormalization group (RG) scaling of the
theory is that of the top quark, $\lambda _t$.
In the large $\tan\beta$ region there will be additional effects.
The one loop RGE of the MSSM,
including $\lambda_t$ effects, is well known~\cite{5,6}.
For our purposes the most important effect is the reduction of the scalar
masses of
$Q_3$ and $U^c_3$
beneath that of the other squarks.
This lightness of the $\tilde{t}_{L}$, $\tilde{b}_L$ and
$\tilde{t}_R$ squarks is very well-known; it is a feature which appears in
the radiative breaking of ${\rm SU}(2) \otimes {\rm U}(1)$ which occurs in
this theory.
When the left-handed down quarks are rotated by the matrix $\mb{V}$ to
diagonalize the quark mass matrix, the non-degeneracy of $\tilde{b}_L$
with $\tilde{s}_L$ and $\tilde{d}_L$
implies that if this rotation is also performed on the squarks they will
acquire an off diagonal mass matrix.
In this paper we work in a mass basis for the squarks, so that the rotation
$\mb{V}$ is done only on the $d_L$ fermions not on the $\tilde{d}_L$
scalars.
This results in the appearance of the KM matrix at the neutralino gauge
vertices.
In particular, for the gluino $\tilde{g}$ we find
$$
{\cal{L}}_{\rm MSSM} \supset \sqrt{2} g_3 (\tilde{d}_L^* T^a\mb{V} d_L)
\tilde{g}^a.\eqno(2)
$$
The phenomenological effects of this flavor mixing at the gaugino
vertex are known to be slight.
There are gluino mediated box diagram contributions to $K^0\bar{K}^0$
and $B^0\bar{B}^0$  mixing.
The contribution to $\Delta m_K$ is negligible, while that to $\varepsilon_K$
and $\Delta m_B$ is less than 10\% of the SM contribution~\cite{5,6}.
Such precise statements are possible because the mixing matrix appearing in
(2) is the KM matrix, and because we know that the gluino and squark masses
are larger
than 150 GeV in the MSSM.
Because the mixing matrix introduces no new phases, the extra contribution to
$B^0\bar{B}^0$ mixing does not effect CP violation in $B$ meson
decays~\cite{BiGi,CLED}. The asymmetries for $B_d \to \pi^+\pi^-$,
$B_d \to \psi K_s$ and $B_s \to \rho K_s$ are
 proportional to $\sin 2\hat{\alpha}$, $\sin 2\hat{\beta}$ and
$\sin 2\hat{\gamma}$
where, as in the SM,
$\hat{\alpha}$, $\hat{\beta}$ and $\hat{\gamma}$ are the angles of the
unitarity triangle which
closes: $\hat{\alpha} + \hat{\beta} + \hat{\gamma} = \pi$.

The superpenguin contribution to $\varepsilon_K'/\varepsilon_K$
is less than about $5 \times 10^{-4}$~\cite{9} and,
given the theoretical uncertainties, is unlikely to be distinguished
from the SM penguin contribution.
In the MSSM a significant flavor changing effect is in the process
$b\to s\gamma$~\cite{6,BiGi,10}.
The recent experimental results from CLEO show that
${\rm B.R.}(b\to s\gamma)$ is
in the
range $(1\div 4) \cdot 10^{-4}$, at 95\% confidence level.
For $m_t = 175 \pm 15$ GeV the SM prediction is
${\rm B.R.}(b \to s\gamma) = (2.9 \pm 1.0)\cdot 10^{-4}$.
These results provide a considerable limit to the MSSM.
However since the MSSM also involves a charged Higgs loop contribution,
the limit does not apply directly to the gluino
loop contribution, which involves the vertex of equation~(2).

The Yukawa interactions for the minimal supersymmetric~SU(5) theory are
given by
$$
W_{\rm SU(5)} = T \bar{\mb{\lambda}}_U TH +
T \mb{\lambda}_D \bar{F}\bar{H} \eqno (3)
$$
where $T$ and $\bar{F}$ are 10 and $\bar{5}$ representations of matter, $H$
and $\bar{H}$ are 5 and $\bar{5}$ Higgs supermultiplets, and the down Yukawa
matrix can be taken to have the form
$\mb{\lambda}_D = \mb{PV}^*\bar{\mb{\lambda}}_D$.
${\mb V}$ is the KM matrix, $\mb{P}$ is a diagonal phase matrix with two
physical phases and
$\bar{\mb{\lambda}}_{U,D}$ are real and diagonal.
Beneath $M_{\rm G}$ phase rotations can be performed so that
$\mb{P}$ does not appear in the low energy interactions.
The Yukawa interactions become those of the MSSM of equation (1)
for the quarks, as well as
$E^c{\mb{\lambda}}_ELH$, for the leptons, with
$\mb{\lambda}_D = \mb{V}^* \bar{\mb{\lambda}}_D$ and
$\mb{\lambda}_E = \mb{V}_{\rm G}^*{\bar{\mb{\lambda}}}_E$, where $\mb{V}$
is the running KM matrix and $\mb{V}_{\rm G}$ its value at $M_{\rm G}$.
For a given $\lambda_t$ the scalar non-degeneracy for
$\tilde{t}_L, \tilde{b}_L$ and $\tilde{t}_R$
are larger than in the MSSM.
This is due to the modified numerical coefficients in the RGE above $M_{\rm
G}$.
More importantly, since $\tau_R$ is unified with the top quark, the
$\tilde{\tau}_R$
has a mass which is lowered compared to that of $\tilde{e}_R$ and
$\tilde{\mu}_R$.
This means that, in the mass basis for both fermions and scalars,
in addition to neutral gaugino flavor mixing for $d_L$ (as in equation~(2)),
there is also gaugino flavor mixing for $e_R$.
Schematically representing the MSSM flavor mixing in the gauge couplings by
$$
(\bar{u}\mb{V} d),\qquad (\tilde{d}^* \mb{V} d)\eqno(4)
$$
that for the minimal~SU(5) theory can be written
$$
(\bar{u} \mb{V} d),\qquad (\tilde{d}^* \mb{V} d),\qquad (\tilde{e}^{c*}
\mb{V}_{\rm G} e^c)\eqno(5)
$$
where all fermion fields are left-handed.

In~SO(10) theories an entire generation is represented by a single spinor: 16.
The Yukawa interaction 16 ${\mb{\lambda}} 16 \Phi$, where $\Phi$ is a 10
dimensional Higgs multiplet, gives mass to the all the fermions, but does not
allow generation mixing.
We consider a minimal~SO(10) model~\cite{DH} with Yukawa interactions which
can be put in the form
$$
W_{\rm SO(10)} = 16 \bar{\mb{\lambda}}_U 16 \Phi_U + 16 \mb{\lambda}_D 16
\Phi_D.\eqno(6)
$$
All scalars of the third generation are split in mass from those of lighter
generations,
so that flavor mixing matrices appear at all neutral gaugino vertices,
except those of the up sector.
Beneath $M_{\rm G}$ the Yukawa interactions have the form
$$
W'_{\rm SO(10)} = Q{\bar{\mb{\lambda}}}_U U^cH_2+
Q \mb{V}^* \bar{\mb{\lambda}}_D \mb{P}^{*2} \mb{V}^\dagger D^c H_1+
E^c \mb{V}_{\rm G}^* \bar{\mb{\lambda}}_E \mb{P}^{*2}
\mb{V}^\dagger_{\rm G} L H_1
\eqno(7)$$
where an asymmetric basis between left and right has been chosen such that
$\mb{V}$ is the usual KM matrix, and $\mb{P}$ is a diagonal phase matrix with
two independent phases, which we choose as
$$
\mb{P}^2 = \pmatrix{e^{i\hat{\varphi}_d}&0&0\cr
               0&e^{i\hat{\varphi}_s}&0\cr
               0&0&1\cr}.\eqno(8)
$$
Using the schematic notation of equations (4) and (5), the flavor mixing of
the minimal~SO(10) theory has the structure
$$
(\bar{u}_L \mb{V} d_L),\quad
(\tilde{d}^* \mb{V} d),\quad
(\tilde{d}^{c*} \mb{VP}^2 d^c),\quad
(\tilde{e}^{c*}\mb{V}_{\rm G}   e^c),\quad
(\tilde{L}^*\mb{V}_{\rm G} \mb{P}^2 L).\eqno(9)
$$

The flavor mixing structure of the minimal models is summarized by
equations~(4), (5) and~(9), and the phenomenological consequences of these
forms are the subject of Section~4 of this paper.
The effects can be classified into two types:
\begin{itemize}
\item[(A)] $(\tilde{d}^*\mb{V} d)$ effects.
Although the mixing matrix is identical for MSSM and the minimal~SU(5)
and~SO(10) models, the effects in the unified models are amplified because the
modified coefficients in the unified RGE lead to larger non-degeneracies
between
$\tilde{b}_L$ and $\tilde{d}_L/\tilde{s}_L$.
This is, however, not the dominant effect.
\item[(B)] Mixing in the $d_R, e_R$ and $e_L$ sectors. We have explored the
consequences of lepton flavor violation in previous papers~\cite{BH,BHS} and
found the signals for $\mu\to e\gamma$ and $\mu \to e$ conversion to be of
great interest, especially in~SO(10) where the mixing in both helicities
implies that amplitudes for the processes can be proportional to $m_\tau$
rather than to $m_\mu$.
Also in the~SO(10) case there are important contributions to the electron and
neutron electric dipole moments, which, in a standard basis and notation for
the~KM matrix, are proportional to
$\sin (\hat{\varphi}_d-2\hat{\beta})$~\cite{DH,BHS}.
In this paper we compare these signals to the hadronic flavor violating ones.
\end{itemize}

\begin{figure}[p]\setlength{\unitlength}{1in}
\caption[Spettro]{Contour plots of the masses of the third generation scalars
in minimal~SO(10) for $m_{\tilde{e}_R}=300\GeV$ and $\lambda_{t\rm G}=1.25$:
(a) $m_{\tilde{\tau}_R}/m_{\tilde{e}_R}$;
(b) $m_{\tilde{b}_L}$;
(c) $m_{\tilde{b}_R}/m_{\tilde{d}_R}$ and
(d) the lightest stop for $\mu<0$.}
\end{figure}

\section{The Scalar Spectrum}
The masses of the scalars of the third generation receive important radiative
corrections from the large $\lambda_t$ coupling in~SU(5) and~SO(10) theories.
The resulting spectrum provides an important signature
of unification, which we present in this section.

In the minimal models there are 6 parameters which play a fundamental role in
determining the spectrum,
flavor and CP violating signals discussed in this paper.
In more general models other parameters may enter,
and we discuss this in Section 6.
The 6 parameters are $\lambda_t$ (the top quark coupling), $m_0$
(the common scalar mass at $M_{\rm Pl}$), $M_0$ (the common gaugino mass at
$M_{\rm Pl}$),
$A_0$ (the common coefficient of the supersymmetry breaking tri-scalar
interactions at $M_{\rm Pl}$),
$B$ (the coefficient of the Higgs boson coupling $h_1h_2$ at low energies)
and $\mu$ (the supersymmetric Higgsino mass parameter).
The solutions of the RGE for the MSSM, minimal~SU(5)
and minimal~SO(10) models has been given previously, including all one loop
$\lambda_t$ effects~\cite{BHS}.
We do not repeat that analysis here, but rather recall the strategy
which we take to deal with this large parameter space
\begin{itemize}
\item[$\lambda_{t\rm G}$] for our purposes it is most useful to parameterize
the top Yukawa coupling by its value at the unification scale
$\lambda_{t\rm G} = \lambda_t (M_{\rm G})$.
This is because the large radiative effects which generate our signals
are induced by the top quark coupling in the unified theory.
Now that the top quark has been found, it may be argued that
$\lambda_{t\rm G}$ should be given in terms of other parameter
$\lambda_{t\rm G} = \lambda_{t\rm G} (m_t, \tan\beta, \alpha_3)$,
where $\tan\beta = v_2/v_1$ is the ratio of Higgs vacuum expectation values.
In fact, for low values of $\tan\beta$, $\lambda_{t\rm G}$ has a strong
dependence on $\alpha_3$, and hence we prefer to keep $\lambda_{t\rm G}$
as the independent parameter.
For larger values of $\tan\beta$, for example $\tan\beta\approx 10$,
and with $m_t = 175 \pm 15\GeV$, $\lambda_{t\rm G}$
cannot be larger than unity.
However, the prediction for $m_b/m_\tau$ requires a larger value
of $\lambda_{t\rm G}$, and hence we
will not consider these larger values of $\tan\beta$ in this paper.
Much larger values of $\tan\beta$, comparable to $m_t/m_b$, do allow
large $\lambda_{t\rm G}$, but in this case there will be many extra
important renormalizations induced by the large coupling $\lambda_b$,
which we have not included.
Hence this paper does not consider the $\tan\beta \approx m_t/m_b$ case.
\item[$m_0$] is traded for the mass of the right hand scalar
electron $m_{\tilde{e}_R}$,  since this is of more physical interest.
\item[$M_0$] is traded for the low energy~SU(2) gaugino mass parameter
$M_2$.
Note that while $M_0/m_0$ may be taken arbitrarily large, this is not
true for $M_2/m_{\tilde{e}_R}$, which is restricted to be less than about
unity.
This is because a large value of $M_0$ generates large scalar masses through
renormalization, especially in the unified theory where Casimirs are
large~\cite{BHS} (we are insisting on $m_0^2>0$).
\item[$A_0$]  is traded for $A_e$, where the selectron
trilinear scalar coupling is $A_e\lambda_e \tilde{L}_e\tilde{e}^c h_1$.
The dimensionless parameter $A_e/m_{\tilde{e}_R}$ is restricted to be in the
range
$-3$ to $+3$ for reasons of vacuum stability.
\item[$B$]  appears in the Higgs potential. On minimizing this
potential, $B$ is traded for $\tan\beta$.
\item[$\mu$]  appears in the Higgs potential. When this
potential is minimized, $\mu^2$ is determined
by $M^2_Z$.
\end{itemize}

Hence the relevant parameter space is
$\{\lambda_{t\rm G}, m_{\tilde{e}_R}, M_2, A_e, \tan\beta,{\rm sign}(\mu)\}$.
All our signals are displayed in the $\{M_2,A_e/m_{\tilde{e}_R}\}$
plane, where $M_2$ and $A_e/m_{\tilde{e}_R}$ are allowed to
run over their entire range.
These planes are shown for representative choices of
$\{ \lambda_{t\rm G}, m_{\tilde{e}_R}, \tan\beta\}$ and negative $\mu$.
Our conclusions do not depend on the sign of $\mu$.

How large are the non-degeneracies amongst the scalars induced
by the coupling $\lambda_{t\rm G}$ in the unified theory?
A simple guess would be that the fractional breaking of degeneracies would be
$ \approx \lambda^2_{t\rm G}/16\pi^2 \ln (M_{\rm Pl}/M_{\rm G})$,
which is a few percent for
$\lambda^2_{t\rm G}=2$.
In fact, the unified theory leads to a large Casimir, and also $\lambda_{t\rm
G}$ may get larger above $M_{\rm G}$, resulting in non-degeneracies which are
an order of magnitude larger than this simple guess.

Numerical results are shown in Figure~1 for the case of
$m_{\tilde{e}_R} = 300\GeV$ in the minimal~SO(10) theory.
The results are insensitive to $\tan\beta$ and to the sign of $\mu$.
There is a large sensitivity to $\lambda_{t\rm G}$.
We take $\lambda_{t\rm G} = 1.25$, which is below the fixed point value
implied by the running of the Yukawa coupling from
$M_{\rm G}$ to $M_{\rm Pl}$~\cite{BHS}.
Figures~1a and 1b, with relatively minor modifications, apply also to the
minimal~SU(5) case with $\lambda_{t\rm G}=1.4$.
Over roughly half of the $A_e/M_2$ plane, the fractional non-degeneracies
are above 30\%.
The fractional non-degeneracy is larger for the sleptons that for the squarks.
This is because a radiative correction to all squark masses proportional to
the gluino mass tends to restore the squark degeneracy.
We call this the ``{\em gluino-focussing\/}'' effect;
it is especially prominent for large gaugino masses.
In~SO(10) the non-degeneracies
of the left-handed and right-handed squarks are very similar. The same is
true for left and right-handed sleptons. This is the most important
difference between the minimal~SU(5) and~SO(10) models: in the~SU(5) case
the left-handed sleptons are essentially degenerate, as are the right-handed
down squarks.

The distinctive, large scalar non-degeneracies of Figure~1 will provide an
important indication of unification.
A precise measurement of these non-degeneracies will provide an essential
component of the elucidation of the flavor structure of the unified theory.

\begin{figure}[p]\setlength{\unitlength}{1in}
\caption[SO10]{Contour plots
in minimal~SO(10) for $m_{\tilde{e}_R}=300\GeV$, $\lambda_{t\rm G}=1.25$,
$\mu<0$, $\tan\beta=2$,
and maximal CP violating phases (see text)
for
(a) ${\rm B.R.}(\mu\to e\gamma)$;
(b) $d_n$;
(c) $\varepsilon_K$;
(d) $\varepsilon'_K/\varepsilon_K$;
(e) $\Delta m_B$;
(f) ${\rm B.R.}(b\to s\gamma)$.
In the hadronic observables only the gluino
exchange contribution is included.}
\end{figure}

\section{Signals of minimal~SO(10)}
The minimal~SO(10) model has flavor mixing angles at all neutral gaugino
vertices, except those involving the up quark. Furthermore, the weak scale
theory involves two additional phases, $\hat{\varphi}_s$ and $\hat{\varphi}_d$,
beyond those of the MSSM, as can be seen from equations~(8) and~(9).
The presence of flavor mixing at neutral gaugino vertices for both helicities
of $e$ and $d$, together with these extra phases, gives a much richer flavor
structure to the minimal~SO(10) model compared to that of the MSSM or
minimal~SU(5) theory.
In fact, for this general reason, the hadronic signals in minimal~SU(5) are
not especially interesting.
An explicit numerical calculation shows that, although somewhat
larger than the corresponding effects in the MSSM,
the gluino exchange contributions to the hadronic observables,
in~SU(5), do not compete with the leptonic flavor violating signals
and are not considered anymore hereafter.

The strong signals in the lepton sector have been stressed
before~\cite{BH,DH,BHS}, and are briefly recalled here. The process $\mu \to
e\gamma$ is induced by a chirality breaking operator which involves the dipole
moment structure $(\sigma^{\mu\nu}F_{\mu\nu})$. In many theories, for example
the minimal~SU(5) theory, this chirality breaking implies that the amplitude is
proportional to $m_\mu$. However, flavor mixing in supersymmetric theories
breaks chirality, if it occurs in both $e_L$ and $e_R$ sectors, and hence in
the
minimal~SO(10) theory terms in the amplitude for $\mu \to e\gamma$ appear
which are proportional to $m_\tau$.
This gives a large rate for $\mu \to e\gamma$, as illustrated in Figure 2a,
for $\tan\beta=2$, $\lambda_{t\rm G} = 1.25$, $m_{\tilde{e}_R} = 300\GeV$
and $\mu<0$.
Figures~3a and~4a show the $\mu\to e\gamma$ rate with all the same
parameters as in Fig. 2a except for $\lambda_{t\rm G} = 0.85$ (Fig.~3a)
or for a scale $M=2.0\cdot 10^{17}\GeV$ for the universal initial condition on
all scalars and gaugino masses (Fig.~4a).
A similar set of diagrams proportional to $m_\tau$ dominates $d_e$, which is
related to the $\mu \to e\gamma$ branching ratio by a simple formula, valid
over all regions of parameter space
$$
{d_e\over 10^{-27}\,e\cdot{\rm cm}} = 1.3 \sin (\hat{\varphi}_d - 2\hat{\beta}
)
\sqrt{ {{\rm B.R.}(\mu\to e\gamma )\over 10^{-12}}}.\eqno(10)
$$
where the KM matrix elements are taken to be approximately real, except for
$V_{td} = \abs{V_{td}} e^{-i\hat{\beta}}$ and
$V_{ub} = \abs{V_{ub}} e^{-i\hat{\gamma}}$.
With this relation, Figures 2a, 3a, 4a can also be used to
predict $d_e/\sin (\hat{\varphi}_d -2\hat{\beta})$.
We know of no reason why $\hat{\varphi}_d$ should cancel $2\hat{\beta}$, which
comes from the KM matrix, so that we do no expect
$\sin(\hat{\varphi}_d-2\hat{\beta})$ to be much less unity.
The process of  $\mu\to e$ conversion in atoms is induced by two operators:
one is the chirality breaking dipole operator involving
$(\sigma^{\mu\nu}F_{\mu\nu})$, with an amplitude proportional to $m_\tau$,
while the other is the chirality conserving operator
involving $(\gamma^\mu\partial^\nu F_{\mu\nu})$.
The derivative in this operator has a scale of the momentum transfer,
which is set by $m_\mu$, so that these contributions are subdominant.
The dominance of the $(\sigma^{\mu\nu}F_{\mu\nu})$ operator implies
that in titanium the ratio $\Gamma (\mu \to e)/\Gamma(\mu~{\rm capture})$
is~200 times smaller than ${\rm B.R.}(\mu\to e\gamma)$.
This result applies over all regions of parameter space of the minimal~SO(10)
model. In any event, it is simply a reflection of the dominance of the
$\sigma^{\mu\nu}F_{\mu\nu}$ operator,
and hence cannot be construed as a unique signature of~SO(10).
However, the processes $\mu \to e\gamma$, $\mu \to e$ conversion and $d_e$
are very incisive probes of~SO(10) superunification, and in the rest of this
section we compare them with probes in the hadronic sector.

\subsubsection{$\varepsilon_K$ and $d_n$}
The dominant gluino-mediated diagram contributing to the $\Delta S=2$
effective Lagrangian involves the exchange of one $\tilde{d}_L$ type squark
and one $\tilde{d}_R$ type squark.
In the limit of keeping only the $\tilde{b}$
contribution, and setting $m_{\tilde{b}_L} = m_{\tilde{b}_R} = M_3$,
this diagram gives:
$$
{\cal{L}}_{\rm eff}^{\Delta S=2} =
{\alpha^2_3(M_3)\over 12 M_3^2} \abs{V_{ts}V_{td}}^2
e^{i(\hat{\varphi}_d-\hat{\varphi}_s)} y^2
[2 (\bar{d}_R^a s^b_L)(\bar{d}^b_L s^a_R)-
 6 (\bar{d}^a_R s^a_L)(\bar{d}^b_L s^b_R) ]\eqno(11)$$
where color indices $a,b$ are shown explicitly.
The parameter $y \approx 0.77$
appears because two of the flavor mixing matrices
are right-handed, and
$$
(V_{G})_{ti}= y V_{ti} \eqno(12)
$$
where $i=d,s$.
This $LR$ contribution is larger than the $LL$ and $RR$ contributions by
about an order of
magnitude, due to the $(m_K/m_s)^2$ enhancement of the hadronic matrix element.
Such an effect is characteristic of~SO(10), since it is not there in the~MSSM
or in
minimal~SU(5).
We use the vacuum insertion approximation:
$$
 \langle K^0 | (\bar{d}_R^a s^a_L)(\bar{d}_L^b s^b_R) | \bar{K}^0\rangle =
3\langle K^0 | (\bar{d}^a_R s^b_L)(\bar{d}^b_L s^a_R) | \bar{K}^0\rangle =
{1\over2} \left(\frac{m^2_K f_K}{m_s + m_d}\right)^{\!\!2},
$$
as seen in lattice calculations~\cite{LATT}.
$f_K$ is normalized in such a way that $f_K\simeq 120\,{\rm MeV}$.
Note that here and elsewhere we do {\em not\/} include the QCD corrections,
unless otherwise stated.

The $\Delta S=2$ gluino-mediated amplitude is important
for $\varepsilon_K$ rather than for $\Delta m_K$, and it gives:
$$
\eqalignno{
\abs{\varepsilon_K}^{\tilde{g}}_{\rm SO(10)}& =
{\alpha^2_3(M_3)\over 9\sqrt{2}M_3^2} {f^2_K m^3_K\over(m_s+m_d)^2\Delta m_K}
 y^2 \abs{V_{ts}V_{td}}^2 \sin (\hat{\varphi}_d - \hat{\varphi}_s)=\cr
&\simeq 2.2 \cdot 10^{-2}\, \sin (\hat{\varphi}_d-\hat{\varphi}_s)
\left( {300\GeV\over M_3}\right)^{\!\!2}
\abs{V_{ts}V_{td}\over 4\cdot 10^{-4}}^2
\left( {180\,{\rm MeV}\over m_s + m_d}\right)^2.&(13)\cr}
$$
At first sight equation~(13) would appear to exclude
colored superpartners less than about 1~TeV; however our simple
analytic estimates are considerable overestimates as they neglect the
compensating effects of $\tilde{d}_L$, $\tilde{s}_L$ exchange, and they do not
give the full dependence on the superpartner parameter space.
Nevertheless, the importance of
$\abs{\varepsilon_K}^{\tilde{g}}_{\rm SO(10)}$ is borne
out by the numerical results, which we discuss shortly.

The two most powerful hadronic probes of the minimal~SO(10) model are
$\varepsilon_K$ and $d_n$, hence we now give our analytic results for $d_n$
which we take to be ${4\over 3} d_d$, where
${\cal{L}}_{\rm eff} =
\frac{1}{2} d_d \cdot \bar{d}\sigma^{\mu\nu} F_{\mu\nu} i\gamma_5d$
and
$$
d_d = e {\alpha_3(M_3)\over 54\pi M_3^2} m_b(M_3) y\abs{V_{td}}^2
{A_b+\mu\tan\beta\over M_3}
\sin(\hat{\varphi}_d -2\hat{\beta})\eqno(14)
$$
where $y$ is given in equation~(12) and we use, as before, the analytic
approximation
of keeping only the gluino diagram with internal $\tilde{b}$ squark, and
set  $m_{\tilde{b}_L} = m_{\tilde{b}_R} = M_3$.
The parameter $m_b(M_3)$ is the running $b$ quark mass
renormalized at $M_3$.
This gives
$$
d_n = 4.2 \times 10^{-26}\, e\cdot{\rm cm}\times \frac{m_b(M_3)}{2.7\GeV}
\abs{ {V_{td}\over 0.01}}^2 {y\over 0.77}
\left({300\GeV\over M_3}\right)^2 \frac{A_b + \mu \tan\beta}{M_3}
\sin(\hat{\varphi}_d -2\hat{\beta}). \eqno(15)
$$
In Figure~2b we show the numerical contour plot for
$|d_n/\sin (\hat{\varphi}_d-2\hat{\beta})|$, and in Figure~2c a contour plot of
$|\varepsilon_K|^{\tilde{g}}_{\rm SO(10)}/
\sin(\hat{\varphi}_d-\hat{\varphi}_s)|$,
where $\varepsilon_K|^{\tilde{g}}_{\rm SO(10)}$
is the contribution to $\varepsilon_K$ from the gluino box diagram only. The
roughly vertical contours, at least in $\varepsilon_K$, reflect the structure
imposed on the scalar non-degeneracy by gluino focussing, shown in Figures~1b
and~1c. This is in marked contrast to the lepton signals of
$\mu \to e\gamma$ and $d_e$ shown in Figure~2a, which reflect the slepton
non-degeneracy of Figure~1a.
Figures~2a, 2b, 2c clearly show that a large $\lambda_{t\rm G}$, as suggested
by $b-\tau$ unification, with the running of the RGE in the full range from
$M_{\rm Pl}$ to $M_{\rm G}$, leads to $\mu\to e\gamma$ as the dominant probe
of the minimal~SO(10) model.
Already the present bound of $5\cdot 10^{-11}$ on the rate excludes a
large portion of the parameter space.
Outside this range, both the gluino exchange contribution to $d_n$ and
$\varepsilon_K$ are, anyhow, negligibly small.
The situation does change, however, if one looks at Fig.s 3 and 4.
As noticed in the previous section, the gluino focussing effect makes
non-degeneracy in the squark sector less prominent than in the slepton sector
and, as such, also less sensitive to a reduction in
$\lambda_{t\rm G}$ and/or in the scale for the initial condition of the RGE.
In turn, although $\mu\to e\gamma$ remains as a very sensitive probe,
it is now possible that gluino mediated contributions to $d_n$ and
$\varepsilon_K$ become relevant, with gluinos in the $(200\div 300)\GeV$
mass range.

For superpartner parameters such that
$\abs{\varepsilon_K}^{\tilde{g}}_{\rm SO(10)} = \varepsilon_K = 2\cdot10^{-3}$,
and for equal phases:
$\hat{\varphi}_d - \hat{\varphi}_s$ =  $\hat{\varphi}_d -2\hat{\beta}$,
$d_n$ is predicted to be very close to its present experimental limit.
Hence $\varepsilon_K$ and $d_n$ provide roughly
comparable probes of this new physics.
However, the new physics in $\varepsilon_K$
must be disentangled from the SM background.

A crucial point emerges from Figures~3c, 4c. For a given
$\hat{\varphi}_d - \hat{\varphi}_s$ it is only over a relatively small region
of the plane that
$\varepsilon_K|^{\tilde{g}}_{\rm SO(10)}$ will make a contribution to
$\varepsilon_K$ that we can disentangle from the SM contribution.
The same statement applies to the planes drawn for different values of
$\{\lambda_{t\rm G}, m_{\tilde{e}_R},\tan\beta, {\rm sign}\mu\}$.
This is partly due to the gluino focussing effect on the scalar masses, but
is also because the SM involves $B_K$, $V_{td}$, $m_t$ in such a
way that it will be very hard to identify contributions which are at the
level of $\varepsilon_K/5$ or less.
Contrast this to the situation with $d_n$, where each factor of~10
improvement in the experimental limit rules out large areas of parameter space.
For this reason we view $d_n$ as an excellent probe of the~SO(10) model.
It has a dependence on the superpartner parameters which is somewhat
orthogonal to that of $d_e$,
as can be seen by comparing Figures~3a and~4a with~3b and~4b.

The neutron electric dipole induced by the~KM phase in the MSSM has been
recently studied in ref.~\cite{dnJap} and is found to be below $10^{-27}
e\cdot{\rm cm}$. In the approximation of neglecting all Yukawa couplings except
the top one in the RGEs, as done here, there is no one loop contribution to
$d_n$ in the~MSSM as in minimal~SU(5).

\subsubsection{$\varepsilon'_K/\varepsilon_K$}
Much present experimental effort is aimed at determining the size of~CP
violation in the direct decays of neutral $K$ mesons:
$\varepsilon_K'/\varepsilon_K$.
How large are the gluino-mediated penguin contributions to this?
The SM contribution is dominated by  $W$ exchange
generation of the penguin operator
$\bar{d}\gamma^\mu s \; \partial^\nu G_{\mu\nu}$,
where $G_{\mu\nu}$ is the gluon field strength, with coefficient
$\propto {\rm Im} V_{ts}V_{td}^*/M_W^2$.
In either the MSSM or minimal~SU(5) or~SO(10) models,
the gluino-mediated penguin contribution does not compete because $M_W$ is
replaced with a larger superpartner mass $m_{\tilde{q}}$ or $M_3$.

However, an interesting new possibility emerges in the minimal~SO(10)
model: a contribution to $\varepsilon_K'/\varepsilon_K$ from a gluino-mediated
 chromoelectric dipole moment operator proportional to $m_b$.
The relevant $\Delta S=1$ effective Lagrangian is, with our usual analytic
assumptions:
$$
\eqalignno{
{\cal{L}}_{\rm eff}^{\Delta S=1} &= g_3(\Lambda_{\rm QCD})
{\alpha_3(M_3)\over36\pi M_3^2} {A_b + \mu \tan\beta \over M_3}
m_b (M_3) y\abs{ V_{ts}V_{td}}\times\cr
&~\times\left\{ e^{i(\hat{\varphi}_d-\hat{\beta})}
  \bar{d}_R\sigma^{\mu\nu} {\lambda\over 2}^a
  s_L G^a_{\mu\nu} + e^{i(\hat{\beta} - \hat{\varphi}_s)}
   \bar{d}_L \sigma^{\mu\nu}
  {\lambda\over 2}^a s_R G^a_{\mu\nu}\right\}.&(16)\cr}
$$
No exact proportionality relation holds between $d_n$ and
$\varepsilon'_K/\varepsilon_K$ since the photon is attached only to the
internal squark line, whereas the gluon, in the chromoelectric dipole
moment, may also be attached to the gluino line.

To evaluate~(16) we use matrix elements~\cite{EEps}
$$
\eqalignno{
\langle \pi\pi, I =0 |
g_s\bar{d}_R\sigma^{\mu\nu} {\lambda\over 2}^a s_L G^a_{\mu\nu}| K\rangle
&=- \langle \pi\pi, I=0 | g_s\bar{d}_L\sigma^{\mu\nu}
{\lambda\over 2}^a s_R G^a_{\mu\nu} | K\rangle =\cr
&=\sqrt{3} {11\over 8} {f^2_K\over f^3_\pi}
{m^2_K\over m_s} m^2_\pi D  \approx 0.37\GeV^2,\cr}
$$
where $D = m^2_K/\Lambda_{\rm QCD}^2 \approx 0.3$, giving
$$
{|\varepsilon_K'|^{\tilde{g}}_{\rm SO(10)}\over|\varepsilon_K |} =
{w\,|{\rm Im} \langle {\cal{L}}_{\rm eff}^{\Delta S =1}\rangle|\over
\sqrt{2}\abs{\varepsilon_K}{\rm Re} A_0}
= 3.1 \times 10^{-4} \left({300\GeV\over M_3}\right)^{\!\!2}
\frac{A_b + \mu \tan\beta}{M_3}
{\sin(\hat{\varphi}_d-\hat{\beta})+\sin(\hat{\varphi}_s-\hat{\beta})\over 2}.
\eqno (17)
$$
We have used $w = 1/22$, ${\rm Re}A_0 = 3.3 \cdot 10^{-7}\GeV$,
$\abs{\varepsilon_K} = 2.3 \times 10^{-3}$ and
$m_b(M_3) = 2.7\GeV$.
This is to be compared with the expectation from the SM for
$m_t = (175 \pm 15)\GeV$:
$\varepsilon_K'/\varepsilon_K = (3\div 10) \cdot 10^{-4}$~\cite{EEps}.

The numerical results for $\varepsilon_K'/\varepsilon_K$ are shown in
Figures~2d, 3d, 4d for $[\sin(\hat{\varphi}_d -\hat{\beta}) +
\sin (\hat{\varphi}_s-\hat{\beta})] = 2$.
Comparing Figures~b for $d_n$ and~d for $\varepsilon_K'/\varepsilon_K$ one
finds that, in the region where these predictions could be of experimental
interest, there is an approximate numerical relation
$$
\left|{\varepsilon_K'\over \varepsilon_K}\right|^{\tilde{g}}_{\rm SO(10)}
\simeq 10^{-4}\left[{
 \sin(\hat{\varphi}_d -\hat{\beta}) +
 \sin(\hat{\varphi}_s - \hat{\beta})\over
2\sin(\hat{\varphi}_d - 2 \hat{\beta})}\right] \times
{d_n\over  10^{-26} e\cdot {\rm cm}}.\eqno(18)
$$
Hence we see that, for the phase ratio in square brackets of unity, the
gluino-mediated contribution to $\varepsilon_K'/\varepsilon_K$ is already
constrained to be not greater than the SM contribution.
Given the theoretical uncertainties in both the penguin and the chromoelectric
dipole matrix elements, we find it unlikely that the gluino-mediated
contribution to $\varepsilon_K'/\varepsilon_K$ could be identified in this
case.

\subsubsection{$\Delta m_{B_d}$}
The rest of this section is devoted to a discussion of $B$ meson signatures of
the minimal~SO(10) model. The gluino-mediated box diagrams for neutral $B$
meson mixing induce an effective Lagrangian
$$
\eqalignno{
{\cal{L}}^{\Delta B=2}_{\rm eff} =& {\alpha^2_3(M_3)\over 12 M_3^2}
\abs{V_{td}}^2
     \bigg\{ e^{2i\hat{\beta}}(\bar{d}_L\gamma^\mu b_L)^2 + y^2
     e^{2i(\hat{\varphi}_d-\hat{\beta})}(\bar{d}_R\gamma^\mu b_R)^2 \cr
&+ ye^{i\hat{\varphi}_d} [2(\bar{d}^i_R b^j_L)(\bar{d}^j_L b^i_R)-
6(\bar{d}^i_R b^i_L)(\bar{d}^j_L b^j_R)]\bigg\}
 + (d\to s, \hat{\beta} \to 0).&(19)\cr}
$$
Using the vacuum insertion approximation, this leads to a contribution to the
mass difference for the neutral $B_d$ mesons of
$$
\Delta m_{B_d}|^{\tilde{g}}_{\rm SO(10)} = {2\alpha^2_3(M_3)\over 9M_3^2}
\abs{V_{td}}^2 f^2_{B_d}m_B\abs{\frac{1}{4} e^{2i\hat{\beta}} +
\frac{y^2}{4} e^{2i(\hat{\varphi}_d-\hat{\beta})}
+ y e^{i\hat{\varphi}_d}}\eqno(20)
$$
where the three terms correspond to $LL$, $RR$ and $LR$
contributions respectively.
For $K^0\bar{K}^0$ mixing the $LR$ terms dominate because of a factor
of $m^2_K/m^2_s$ enhancement of the matrix element.
No such factor occurs in the $B$ system, but the vacuum insertion
approximation suggests that the $LR$ term still dominates, giving
$$
\Delta m_{B_d}|^{\tilde{g}}_{\rm SO(10)} \simeq 2.7\cdot 10^{-10}\,{\rm MeV}
\left({300\GeV\over M_3}\right)^{\!\!2}
\left({f_B\over 140\,{\rm MeV}}\right )^{\!\!2},\eqno(21)
$$
with $f_B$ normalized in the same way as $f_K$.
In the limit that the $LR$ operator contributions dominate both
$\varepsilon_K|^{\tilde{g}}_{\rm SO(10)}$ and
$\Delta m_{B_d}|^{\tilde{g}}_{\rm SO(10)}$,
we can write a relation
$${\abs{\varepsilon_K}^{\tilde{g}}_{\rm SO(10)}\over
\Delta m_{B_d}|^{\tilde{g}}_{\rm SO(10)}} \simeq
{1\over 2\sqrt{2}}{f^2_K\over f^2_B} {m^3_K\over \Delta m_K (m_s +m_d)^2 m_B}
y\abs{V_{ts}}^2 \sin (\hat{\varphi}_d -\hat{\varphi}_s)\eqno(22)
$$
which is approximately independent of the superpartner spectrum and of
$\abs{V_{td}}$. Inserting numbers:
$$
{\Delta m_{B_d}|^{\tilde{g}}_{\rm SO(10)}\over
3.5 \cdot 10^{-10}\,{\rm MeV}}\simeq
{0.1\over \sin(\hat{\varphi}_d-\hat{\varphi}_s)}
{ \abs{\varepsilon_K}^{\tilde{g}}_{\rm SO(10)}\over 2.3 \times 10^{-3}}
\times\left( {f_B\over 140\,{\rm MeV}}\right)^2
 \left( {m_s + m_d \over 0.18\GeV}\right)^2
 \abs{ {0.04\over V_{ts}}}^2\eqno(23)
$$
demonstrating that $\Delta m_{B_d}|^{\tilde{g}}_{\rm SO(10)}$ can only be a
large fraction of the observed $\Delta m_{B_d}$ if
$\sin(\hat{\varphi}_d - \hat{\varphi}_s)$ is small, unless the vacuum
insertion approximation for the $LR$ operator is an overstimate.

The numerical results for $\Delta m_{B_d}|^{\tilde{g}}_{\rm SO(10)}$,
assuming dominance of the $LR$ contribution, are shown as a contour
plot in Figures 2e, 3e, 4e.

The contours of Figures~2e, 3e, 4e are normalized to the observed value
$\Delta m_{B_d} = 3.5\cdot 10^{-10}\,{\rm MeV}$.
As in the comparison of $d_n$ and $\varepsilon_K$ with the leptonic signal,
for values of the parameters as in Fig.~2, $\Delta m_{B_d}|^{\tilde{g}}$
is constrained to be too small to be of interest.
We therefore consider only the cases of Fig.s~3, 4.
A useful parameter in our discussion of the phenomenology is
$$
r = {\Delta m_{B_d}|^{\tilde{g}}_{\rm SO(10)}\over \Delta m_{B_d}|_{\rm SM}},
\eqno(24)
$$
with the top mass in the SM contribution set to $175\GeV$.
In particular, it is convenient to consider three regions of the supersymmetric
parameter space: A, B and C:
\begin{itemize}
\item[A]
$r \ll 0.1$. In this region we find that all gluino-mediated contributions to
the hadronic observables provide only very small deviations from the~SM
predictions.
The only exception to this is $d_n$.
{}From Figures 3e and 4e we see that this is a very large region.
\item[B]
$ r \approx 0.1.$ A point with $r=0.1$ is provided by:
$\lambda_{t\rm G} = 0.85$,
$\tan\beta=2$, $m_{\tilde{e}_R} = 300\GeV$, $A_e/m_{\tilde{e}_R}=2$
and $M_2=80\GeV$.
At this point, $M_3 = 250\GeV$, $m_{\tilde{b}} = 200\GeV$,
$m_{\tilde{q}}=400\GeV$ and $m_{\tilde{t}_1} =100\GeV$.
This illustrates that region~B can be reached without taking superpartner
masses too close to their present lower limits.
\item[C] $r \approx 0.5$. An example of a point in this region
is provided by: $\lambda_{t\rm G} = 0.85$, $\tan\beta=2$,
$m_{\tilde{e}_R} = 300\GeV$,
$A_e/m_{\tilde{e}_R}=1$ and $M_2 = 50\GeV$.
At this point, other masses are approximately: $M_3 = 150\GeV$,
$m_{\tilde{b}} = 150\GeV$,
$m_{\tilde{q}} = 300\GeV$ and $m_{\tilde{t}} =100\GeV$.
The gluino mass is now below $200\GeV$,
so we expect that this region will be probed at the Fermilab collider.
It is clear that values of $r$ larger than about 1 are excluded by present
limits on the gluino mass.
\end{itemize}
The majority of our discussion will concern $0.05 < r < 1$ (which includes
regions B and C) as this is the region where the hadronic signatures are
important. However, it is important to realize that much of the parameter
space has $r \ll 0.1$, and hence can only be probed by the lepton signals.

To discuss the phenomenology of these parameter regions, it is important to
consider the theoretical predictions for $\abs{\varepsilon_K}$ and for
$\Delta m_{B_d}$, which include both SM and gluino mediated
contributions (neglecting other supersymmetric contributions).
We find a useful approximation to be:
$$
\abs{\varepsilon_K} \approx 2.26 \times 10^{-3} {\eta B_K\over 0.5}
\abs{{V_{td}\over 0.01}}^2
[1.8 \sin 2\hat{\beta} +
11.5 r \sin (\hat{\varphi}_d-\hat{\varphi}_s)]\eqno(25)
$$
and
$$
\Delta m_{B_d} \simeq 2.1 \times 10^{-10}\,{\rm MeV}
{\eta B_B\over 0.5}
\left( {f_B\over 140\,{\rm MeV}}\right)^2
\abs{{V_{td}\over 0.01}}^2 \abs{ e^{2i\hat{\beta}} + re^{i\hat{\varphi}_d}}
\eqno(26)
$$
where in each equation the first term, involving $\hat{\beta}$, is the SM
result while the second term, involving $r$, is the supersymmetric
contribution.
Note that we have set $m_t = 175\GeV$.
We have also introduced a~QCD correction $\eta$ times a fudge factor
$B_K$ ($B_B$) for the matrix elements of the appropriate operators.

A natural expectation is that all phases, $\hat{\beta}$, $\hat{\gamma}$,
$\hat{\varphi}_d$ and
$\hat{\varphi}_s$, and their differences, are of order unity. This would
exclude
region~C as $\abs{\varepsilon_K}$ is predicted to be too large. We will discuss
regions~A and~B when the phases are large. In region~A there is little to say,
the supersymmetric contributions provide small corrections, especially for
$\Delta m_{B_d}$. In region~B supersymmetric contributions to
$\abs{\varepsilon_K}$ are as important as the SM contribution, however the
corrections to $\Delta m_{B_d}$ are small. Fits to the data will therefore
yield the usual value for
$\abs{V_{td}}$, but $\sin2\hat{\beta}$ will be replaced by
$[\sin 2\hat{\beta} + 7 r
\sin(\hat{\varphi}_d -\hat{\varphi}_s)]$ and will change by a large amount.

In the small region C, $\sin(\hat{\varphi}_d-\hat{\varphi}_s) \ltap 0.1$.
The supersymmetric corrections to $\Delta m_{B_d}$ can be significant,
so that $\abs{V_{td}}$ may change by as much as 50\%.
Fits to data are  now more complicated as they involve
$\hat{\beta}$, $\hat{\varphi}_d$ and $\hat{\varphi}_s$.
Since all phases have the same origin, it is plausible that in region~C they
are all small, of order 0.1. In this case the CP violation which has been
observed in nature is produced dominantly by sources other than the KM matrix.
Although we do not find it likely, the KM matrix could be real in regions~B
and~C.

Figures~3,4b and~3,4d show the behavior of $d_n$ and
$\varepsilon_K'|^{\tilde{g}}_{\rm SO(10)}/\varepsilon_K$ in these regions.
Region~C is clearly excluded by $d_n$ unless $\hat{\varphi}_d-2\hat{\beta}$
is a small
phase, which again suggests that all phases should be small in this region.
In regions~B and~C, $d_n$ is close to discovery.
A search to the level of $10^{-27} e\cdot{\rm cm}$ will probe a substantial
fraction of region~A.
In regions~B and~C, the supersymmetric contribution to
$\varepsilon_K'/\varepsilon_K$ is expected to be at the $10^{-4}$ level.
Whether it can be distinguished from the SM contribution is very dependent on
the sizes of the phases which appear, $\hat{\varphi}_d -\hat{\beta}$ and
$\hat{\varphi}_s - \hat{\beta}$, compared to the phase $\hat{\varphi}_d -
\hat{\varphi}_s$ that occurs in
$\varepsilon_K$.

\subsubsection{$\Delta m_{B_s}$}
The expression for the gluino-mediated contribution to $B_s$ mixing is
obtained from equation~(20) by the replacements:
$V_{td} \to V_{ts}$, $\hat{\beta} \to 0$, $\hat{\varphi}_d \to \hat{\varphi}_s$
and $f_{B_d}\to f_{B_s}$, giving
$$
\Delta m_{B_s}|^{\tilde{g}}_{\rm SO(10)}  = {2\alpha^2_3(M_3)\over 9 M_3^2}
\abs{V_{ts}}^2 f^2_{B_s} m_B \abs{ {1\over 4} + {y^2 \over
4}e^{2i\hat{\varphi}_s} + y e^{i\hat{\varphi}_s}}.\eqno(27)
$$
If the $LR$ contributions dominate $\Delta m_B|^{\tilde{g}}_{\rm SO(10)}$,
we find
$$
{x_s\over x_d} \simeq \abs{ {V_{ts}\over V_{td} }}^2 {f^2_{B_s}\over f^2_{B_d}}
\abs{
{1+ re^{i\hat{\varphi}_s}\over e^{2i\hat{\beta}} + re^{i\hat{\varphi}_d}}}
\eqno(28)
$$
valid for any value of $r$.
Deviations from the SM prediction are $\ll$ 10\%, $\simeq$ 10\%,
$\approx$ 100\% for regions~A, B and~C.

\begin{figure}[p]\setlength{\unitlength}{1in}
\caption{Same as in fig.~2 except for $\lambda_{t\rm G}=0.85$.}
\end{figure}

\begin{figure}[p]\setlength{\unitlength}{1in}
\caption{Same as in fig.~2 except for
the initial conditions on the RGEs taken at $2.0\cdot 10^{17}\GeV$.}
\end{figure}

\subsubsection{CP violations in $B$ decays}
When a tagged neutral $B$ meson decays to~CP eigenstate $a$, there is an
oscillatory term in the decay rate proportional to
$\sin(\phi_M +\phi_a) $ $\sin(\Delta m_{B}\,t)$ which is of opposite sign for
$B^0$ and $\bar{B}^0$ decay and therefore violates~CP.
The phase $\phi_M$ is the phase of the appropriate $B$ meson mixing amplitude
$M_{12}$, while the phase $\phi_a$
is the~CP violating phase of the decay amplitude for $B^0 \to a$.
The values of $\sin(\phi_M + \phi_a)$ for various $a$ in the SM
are shown in the first column of Table~1.

\begin{table}
\begin{center}
\vskip 20pt
\begin{tabular}{|c|c|c|}
\hline
&\vrule height 3.5ex width 0pt depth 2.5ex\parbox[c]{12em}{
\centering{Standard Model\\ and Minimal~SU(5)}} &
\parbox[c]{12em}{\centering{Minimal~SO(10)}}\cr
\hline
$d_e, d_n$& --- &$\sin(\hat{\varphi}_d -2\hat{\beta})$\cr
\hline
$\varepsilon_K$&$\sin2\hat{\beta}$&$\sin(\hat{\varphi}_d -\hat{\varphi}_s)$\cr
\hline
$\varepsilon_K'/\varepsilon_K$&$\sin\hat{\beta}$&
$\sin(\hat{\varphi}_d-\hat{\beta})+ \sin(\hat{\varphi}_s-\hat{\beta})$\cr
\hline
\parbox{6em}{$B_d\to \pi^+\pi^-$}&
$\sin(2\hat{\beta} + 2\hat{\gamma})$&$\sin(\hat{\varphi}_d + 2\hat{\gamma})$\cr
\hline
\parbox{6em}{$B_d \to \psi K_s$}& $\sin2\hat{\beta}$&$\sin\hat{\varphi}_d$\cr
\hline
\parbox{6em}{$B_d\to D^+D^-$}&$\sin2\hat{\beta}$&$\sin\hat{\varphi}_d$\cr
\hline
\parbox{6em}{$B_s\to \rho K_s$}&
$\sin 2\hat{\gamma}$& $\sin(\hat{\varphi}_s +2\hat{\gamma})$\cr
\hline
\parbox{6em}{$B_s\to \psi K_s$} & --- &$\sin\hat{\varphi}_s$\cr
\hline
\end{tabular}
\end{center}

\medskip

\caption{
$\hat{\beta}$ and $\hat{\gamma}$ are defined by:
$V_{td} = \abs{V_{td}} e^{-i\hat{\beta}}$,
$V_{ub} = \abs{V_{ub}} e^{-i\hat{\gamma}}$.
$\hat{\varphi}_{d,s}$ are defined by equations~(8) and~(9).
``---'' indicates signal is too small to be of experimental interest.
For $B$ meson decays: in the Standard Model and minimal~SU(5) theory the
entry
gives the CP violating coefficient of the $\sin \Delta m_{B}t$ oscillatory
term.
For the minimal~SO(10) model the entry gives the contribution to this
coefficient from the gluino exchange contribution to $M_{12}$.
This must be combined with the SM contribution, as shown in
equations~(25) and~(26).}
\end{table}

In supersymmetric theories $\phi_a$ is the same as in the SM:
diagrams involving superpartners provide only very small corrections to $b$
quark decay amplitudes. Hence the possible signals of new physics are via the
mixing amplitude phase $\phi_M$.
In the MSSM and minimal~SU(5) models the supersymmetric contributions to the
$B$ mixing amplitude have the same phase as the SM contribution.
Hence $\phi_M$ is unaltered, and the first column of Table~1 applies also to
the MSSM and minimal~SU(5) theories. However, as can be seen from
equations~(20) and~(28), in the minimal~SO(10) model the supersymmetric
contributions to $B_{d,s}$ mixing have phases
$\simeq \hat{\varphi}_{d,s}$. In the
case that these supersymmetric contributions to $B$ meson mixing dominate
the SM contribution, the quantity $\sin(\phi_M+\phi_a)$, for various
final states $a$, is shown in the 2nd column of Table~1.
This situation or $r \approx 1$ can occur, but over most of
parameter space $r <1$. Since
$$ M^d_{12} \simeq \abs{M^d_{12}}_{\rm SM}
(e^{2i\hat{\beta}} + r\ e^{i\hat{\varphi}_d})\eqno(29a)$$
$$ M^s_{12} \simeq \abs{M^s_{12}}_{\rm SM}
( 1 + r\ e^{i\hat{\varphi}_s})\eqno(29b) $$
and the relevant mixing phase $\phi_{M_i}$ is the phase of $M^i_{12}$,
we find that in regions~A and~B
$$ \phi^d_M \simeq
2\hat{\beta} + r \sin (\hat{\varphi}_d-2\hat{\beta})\eqno(30a)$$
$$ \phi_M^s \simeq r \sin \hat{\varphi}_s\eqno(30b)$$
Hence when $r$ is small the deviations from the SM
pattern of CP violation in neutral $B$ meson decays is proportional to $r$,
and is also small.

In region~C the phases $\phi_M^i$ deviate considerably from the SM form.
For example for $r=1$,
$\{\phi^d_M, \phi^s_M\} =
\{{\hat{\varphi}_d\over 2} + \hat{\beta}, {\hat{\varphi}_s\over 2}\}$,
which differs greatly from  $\{2\hat{\beta}, 0\}$ of the SM.
In this region we have argued that it is likely that all phases are
small, in which case the mixing phases are
$\{2\hat{\beta} +r\hat{\varphi}_d/(1+r), r\hat{\varphi}_s/ (1+r)\}$.
The most notable feature is that, unlike the SM, all asymmetries
should be small.
We stress again that region~C only corresponds to a very small portion of
the parameter space.

\subsubsection{$b\to s\gamma$}
Finally we consider the process $b\to s\gamma$.
The effective Lagrangian for $b \to s\gamma$ can be written in the general
form:
$$
{\cal{L}}_{\rm eff}^{b\to s\gamma} = {e\over 2} m_b(m_b)
[A_L \bar{s}_R \sigma^{\mu\nu} b_L F_{\mu\nu} +
 A_R \bar{s}_L \sigma^{\mu\nu} b_R F_{\mu\nu}]\eqno(31)
$$
in which case the branching ration for $b \to s\gamma$ is given in terms of
the semi-lepton branching ratio via
$$
\eqalignno{
{\rm B.R.}(b\to s\gamma)
 & ={\rm B.R.}(b\to ce\bar{\nu})
    {48 \pi^3\alpha\over G^2_F} {\abs{A_L}^2 + \abs{A_R}^2\over
    \abs{V_{cb}}^2 I (1- {2\over 3\pi} \alpha_3(m_b) f)}\cr
 & = 1.3 \times 10^{13}\GeV^4 (\abs{A_L}^2 + \abs{A_R}^2)&(32)\cr}
$$
where $I\simeq 0.5$ is a phase-space factor and $f\simeq 2.4$
is a QCD correction factor, both occurring in ${\rm B.R.}(b\to ce\bar{\nu})$.

In our usual analytic approximation we have
$$
A^{\tilde{g}}_R = {8\over 27} {\alpha_3(M_3)\over 12\pi M_3^2} \abs{V_{ts}}
\left(-7 + \eta_b {A_b + \mu \tan\beta \over M_3} \right)\eqno(33a)
$$
$$
A_L^{\tilde{g}} = y e^{i\hat{\varphi}_s} A^{\tilde{g}}_R\eqno(33b)
$$
where $\eta_b = m_b(M_3)/m_b(m_b)$.  We therefore obtain
$$
{\rm B.R.}(b\to s\gamma)|^{\tilde{g}}_{\rm SO(10)} = 1.1 \cdot 10^{-4}
\left({300\GeV\over M_3}\right)^4
\left(1 - \eta_b{A_b + \mu \tan\beta\over 7M_3}\right)^2.\eqno(34)
$$
Note that this branching ratio is obtained by simply squaring the gluino
amplitude, and it ignores the SM and charged Higgs contributions,
chargino contributions and their interferences.

The numerical result for the gluino contribution to $b\to s\gamma$ are shown
in Fig. 2f, 3f, 4f. In view of the uncertainties on the SM
contribution to this process, they can hardly play a significant role in any
situation.
The rate for $b\to s\gamma$ is on the other hand known to place a constraint
on the parameter space of the MSSM mostly determined from charged Higgs and
chargino exchanges~\cite{10}.
We notice that in the parameter space displayed in all plots of
Fig.s~1$\div$4 the charged Higgs mass ranges from $300\GeV$ to $1000\GeV$.
Correspondingly only a very small region of the~SO(10) parameter space is
excluded by $b\to s \gamma$,
where the $\mu \to e\gamma$ and $d_e$ signatures can be seen.

In the minimal~SO(10) model the best signatures are the lepton flavor
violating processes and the electric dipole moments of the electron and
neutron. These signatures can be probed by future experiments over a wide
range of parameter space.
Over some of this parameter space gluino-mediated contributions to
$\varepsilon_K$ are significant. Over a  restricted region of parameter space
gluino-mediated contributions to
$\varepsilon_K'/\varepsilon_K$ and to $\Delta m_B$ could be identified. The
latter could lead to deviations from the pattern of~CP violations in neutral
$B$ meson decays expected in the SM. In certain small regions of
parameter space the deviations from the SM could be very large.
However, over most of parameter space, the relative merits of the various
signals are as summarized in Table~2, shown in the conclusions.

\section{The Assumptions.}
The flavor and CP violating signals which we compute are induced by the top
Yukawa coupling of the unified theory. Although the calculations of this
paper are done in specific simple models, the signals occur in any theory
which satisfies three criteria (barring some kind of
flavor symmetry restoration at the unification scale):
\begin{itemize}
\item[i.] At least one helicity of the $\tau$ lepton is unified in the same
representation as the top quark.
\item[ii.]  Supersymmetry is effectively unbroken down to the weak scale.
\item[iii.] The supersymmetry breaking parameters are hard
(have no power-law momentum dependence) at the scale
$M_{\rm G}$ of the unified interactions.
\end{itemize}
It is certainly possible to construct theories without each of these
assumptions.
However, the predominant paradigm of supersymmetric unification
does satisfy all three criteria.
In this section we give arguments in favor of each of these assumptions.

In unified theories with three generations only, it is inevitable that the
first assumption is justified.
In~SU(5) or~SO(10) there must be some lepton in the same irreducible
representation as the top quark.
This could not be dominantly the $e$ or $\mu$,
otherwise the signals that we are discussing, such as $\mu \to e\gamma$,
would be much larger than the present experimental limit.
Hence, to very high accuracy, the top quark is unified with the $\tau$ lepton
in this case.

In unified models with $N+3$ generations and $N$ mirror generations, there is
no fundamental reason why the top quark and $\tau$ need be in the same
representation~\cite{DP}.
The lepton unified with the top quark could be superheavy.
The states of the light generations will be determined by the structure of the
superheavy masses which marry the $N$ mirror generations to $N$ of the
generations.
These mass matrices may break the unified group so that the light states do
not fill out complete representations of the unified group.
Although such rearrangement of generations is possible, it would typically
lead to a Kobayashi-Maskawa matrix with order unity
intergenerational mixing, and hence appears to us not to be preferred.

The second assumption, of weak-scale supersymmetry, is motivated by the
successful prediction of the weak mixing angle, at the percent level, in
superunified models.
Furthermore, the dynamical breaking of the electroweak symmetry induced by the
large top Yukawa coupling connects the scale of supersymmetry breaking to the
$Z$ boson mass.

We believe the third assumption is that which is most open to question.
There is no compelling physical mechanism for supersymmetry breaking.
If the flavor and~CP violating signals are shown to be absent to a high
degree, then it may be a sign that the supersymmetry breaking is soft at
scale $M_{\rm G}$, and is  not convincing evidence that quark-lepton
unification is false.
If the breaking of  supersymmetry is communicated to the particles of the
MSSM at energy scales much less that $M_{\rm G}$, then the supersymmetry
breaking interactions will not reflect any information about the unification
at higher energy, and our signals disappear.
Our signals are present in theories  where supersymmetry breaking occurs in a
hidden sector (with fields $Z_i$) such as can occur in
supergravity~\cite{SuGra}. This sector is called ``hidden'' because beneath
some scale $M$ there are no renormalizable interactions which couple the hidden
fields to those of the MSSM (denoted $\Phi_a$).
Thus beneath $M$ the communication between these sectors is solely via
non-renormalizable operators such as
$M^{-1} [Z_i\Phi_a \Phi_b \Phi_c]_F$,
$M^{-2}[Z^\dagger_iZ_j \Phi^\dagger_a \Phi_b]_D$.
An important assumption is that the physics at scale $M$, which generates
these operators is flavor-blind, treating all generations equally.
Considering the $D$ operator for simplicity,
its coefficient at the scale $M$ can therefore be written as
$\lambda_{ij}\delta_{ab}$.
On renormalizing this operator to lower energies it will receive
radiative corrections from the interactions of both observable and hidden
sectors.
However the hidden sector interactions are flavor-blind, so these
renormalizations
maintain the form $\lambda_{ij}\delta_{ab}$ and simply renormalize
$\lambda_{ij}$.
When supersymmetry breaks in the hidden sector we insert $F_i$ vacuum
expectation values into the operator to generate a supersymmetry breaking mass
for the observable scalar fields $m^2_{ab} = (\lambda_{ij}
F^*_iF_j/M^2)\delta_{ab}$. In the absence of observable sector renormalizations
this is a universal mass. However, the factor $\delta_{ab}$ appeared because of
the flavor independence of the physics at scale $M$ which generated these
non-renormalizable operators.
Beneath $M$, the observable interactions, which do depend on flavor,
renormalize the coefficient away from proportionality to $\delta_{ab}$.
Furthermore, as far as the observable interactions are concerned, it is
simply a question of renormalizing the mass operator
$\Phi^\dagger_a\Phi_b$ from $M$ down to low energies.

This framework is not ideal for two reasons.
Firstly we do not understand why the physics at $M$ which generates these
operators
should be flavor independent.
If it grossly violated flavor symmetry between the lightest two generations, it
would lead to $m_{\tilde{e}}$ and $m_{\tilde{\mu}}$ being very different,
giving ${\rm B.R.}(\mu \to e \gamma) \approx  10^{-4}$.
Hence we simply impose this initial flavor independence as an experimental
necessity.
Secondly, supersymmetry breaking occurs at an intermediate scale,
$F_i^{1/2} \approx (M_W M_{\rm Pl})^{1/2}$,
the origin of which is not understood.

Nevertheless, this framework can occur in the context of $N=1$ supergravity
theories, in which case $M$ is the reduced Planck mass, $M_{\rm Pl}$. So far it
has appeared preferable to alternative schemes with softer supersymmetry
breaking, at least because gravity provides the desired non-renormalizable
interactions.

\section{Conclusions}
In this paper we have studied hadronic flavor and CP violating phenomena
generated by the large top quark coupling in supersymmetric grand unified
theories.
We have computed the gluino-mediated contributions to $\varepsilon_K$,
$\varepsilon_K'/\varepsilon_K$,
$\Delta m_B$, $b\to s \gamma$, $d_n$ and CP violation in neutral $B$
meson decays in two simple models.
The physics at the unified scale, $M_{\rm G}$, is reflected
at low energies in the scalar superpartner spectrum and in flavor mixing
matrices
at neutral gaugino vertices, which have characteristic
forms for the minimal~SU(5) and~SO(10) models.
In the minimal~SU(5) model the flavor mixing matrices  occur at
all neutral gaugino vertices for the $d_L$ and $e_R$ sectors,
while in the minimal~SO(10) model mixing occurs also in the $d_R$ and $e_L$
sectors.

An important, universal, feature of the hadronic signals is that they have a
much larger
dependence on the gaugino mass than the leptonic signals.
A large gluino mass contributes a large flavor-independent radiative
correction to the squark masses, thus reducing the non-degeneracies produced in
the unified theory.
This gluino focussing effect can be seen in Figures~1b,c,d where the squark
mass shows a strong dependence on the gaugino mass. In the lepton
sector the gaugino focussing is much less important, as can be seen from a
comparison of Figures~1a and~1b,c,d.

The hadronic flavor-changing and CP violating effects of the minimal~SU(5)
theory are very similar in nature to those of the MSSM, although
numerically somewhat larger.
The most important limit on the parameter space is
therefore provided by $b\to s\gamma$,
and it is unlikely that the gluino mediated contribution be dominant~\cite{11}.
However, there remain large regions of parameter space where the rare $\mu$
processes, such as $\mu \to e \gamma$, are large and provide the only probe
of this new flavor physics.

The additional flavor mixing matrices of the minimal~SO(10) model make the
hadronic flavor and CP violating signals larger and richer than in the~SU(5)
model, as was also the case for the leptonic channels.
A study of the contour plots of Figures~2,3,4 shows that
a critical role is played by the value of $\lambda_{t\rm G}$ and/or of the
scale $M$ for the initial conditions on the RGEs.
The hadronic flavor and CP violating signals can be significant, relative to
the leptonic ones, only for relatively low values of $\lambda_{t\rm G}$  and/or
$M$. This is an indirect consequence of the gluino focussing effect. In such a
case, even for a not too light gluino, the discovery of $d_n$ may be possible.

As the gluino mass is lowered, with all phases of order unity, the first
process which acquires an important gluino-mediated contribution is
$\varepsilon_K$. Most striking is the possibility that, even with colored
scalars heavier than $300\GeV$, $\varepsilon_K$ may receive non-KM
supersymmetric contributions as large as the SM contribution.
This could be identified by a failure of the SM to accommodate the
observed values of $\varepsilon_K$, $\Delta m_B$ and $\abs{V_{ub}}$.
At present such fits are limited by the $f^2_B$ uncertainty in $\Delta m_B$,
which amounts to a 50\% effect.
In this region, where the supersymmetric contribution to $\varepsilon_K$ is
comparable to the SM one, and where all phases are of order
unity, $\Delta m_B$ receives a correction from gluino-mediated
diagrams at most of $(10\div 20)$\%.
This leads to deviations from the SM pattern of~CP violation in
neutral $B$ meson decay at most of $(10\div20)$\% level.

For still lighter values of the gluino mass, in the region of $200\GeV$, the
gluino mediated contribution to $\varepsilon_K$ is so large that a combination
of phases must be made small.
This suggests that in this region all the CP violating phases are small.
Nevertheless the gluino-mediated contribution to $\Delta m_B$ can be comparable
to that of the SM, meaning that although the CP asymmetries in $B$
meson decay are small they show very large deviations from those predicted by
the SM.
The most salient features of our results are summarized in Table~2.

\begin{table}
$$
\begin{tabular}{|c|c|c|}\hline
&\multicolumn{2}{|c|}{Minimal}\\ \cline{2-3}
&SU(5)&SO(10)\cr \hline\hline
$\mu\to e\gamma, \mu \to e$ &$\surd\surd$&$\surd\surd$\cr \hline
$d_e, d_n$& --- &$\surd\surd$\cr \hline
\vrule height 3.3ex width 0pt depth 2ex
\parbox[c]{7em}{\centering{CP violation\\ in $B^0$ decays}}& ---&$\surd$\cr
\hline\hline
$\varepsilon_K$& --- &**\cr \hline
$\varepsilon_K'/\varepsilon_K$& --- &*\cr \hline
$\Delta m_B$& --- &*\cr \hline
$b\to s\gamma$&*&*\cr \hline
\end{tabular}\hskip 1cm
\parbox[c]{6cm}{
\caption[Summary]{summary of flavor and CP\\ violating signals:}
$$\left\{
\begin{array}{cl}
\surd\surd & \hbox{very important searches}\cr
\surd      & \hbox{significant searches}\cr
\hbox{---} & \hbox{not relevant} \cr
*   & \hbox{constraint on parameters}\cr
**  & \hbox{dominant constraint}
\end{array}\right.$$}
$$
\end{table}

We have chosen to study the minimal~SU(5) and~SO(10) models because the origin
of the flavor violating effects are dominated by the top quark coupling of the
unified theory, and because the flavor mixing matrices are simply related to
the KM matrix. In more general models one expects that
\begin{itemize}
\item The flavor mixing matrices at the gaugino vertices have the same
hierarchical pattern of mixing as the KM matrix, but have entries which
differ numerically from those of the KM matrix.
\item The squark and slepton masses may receive important radiative
corrections to their mass matrices from couplings in the unified theory other
than $\lambda_t$.
\end{itemize}
How will our conclusions be modified for these theories?
The differing flavor mixing matrices increase the uncertainties in the
amplitudes.
Hence, the relative importance of $\varepsilon_K$, $b \to s \gamma$,
$\Delta m _B$ and $\mu \to e \gamma$
may change, causing the contours of Figures~2, 3, 4 to shift by, say,
factors of~3.
This could mean that the modifications to CP violation in $B$ decays are
larger (or smaller) than for the minimal models. The additional radiative
corrections to the scalar mass matrices will similarly increase uncertainties.
Those radiative corrections which produce further non-degeneracies will
enhance our effects, while radiative corrections which produce
flavor-changing scalar masses could add or subtract to our effects, depending
on the signs.
Barring some sort of flavor symmetry restoration at $M_{\rm G}$,
precise cancellations are unlikely, and certainly would not be
expected to occur in more than one process.
Hence we believe that, to within factors of~2 or~3 in amplitude, the results
of this paper can be interpreted as the minimum expected signatures of all
models which satisfy the assumptions discussed in the previous section.

The gluino-focussing effect will be present in all theories. It is unaffected
by changes in the flavor mixing angles, and its effects are enhanced if the
unified theory produces larger squark non-degenerecies than discussed here.
Hence we can state very generally that:
\begin{itemize}
\item[(A)] Hadronic flavor and CP violating processes exclude only very small
regions of parameter space, those with low gluino mass.
\item[(B)] For slightly higher values of the gluino mass, there are very
interesting contributions, especially to $\varepsilon_K$ but also to $\Delta
m_B$, which could be discovered by the failure of SM fits to
these quantities and by future measurements of~CP violation in $B$ decays.
\item[(C)] Lepton flavor violation, such as $\mu \to e\gamma$, and electric
dipole moments, $d_e$ and $d_n$, provide the most powerful probe of this
flavor physics of unified theories. This is because, unlike the hadronic
probes, the signals could be observed over a very wide  region of parameter
space.
\end{itemize}


\end{document}